\documentclass[a4paper,10pt]{aa}
\usepackage[utf8x]{inputenc}
\usepackage{graphicx,amsmath}
\usepackage{txfonts,color,fixltx2e}

\newcommand{\filep}{}
\usepackage{natbib} 
\bibpunct{(}{)}{;}{a}{}{,}
\begin{document}
\title{Variation in sunspot properties between 1999 and 2011 as observed with  the Tenerife Infrared Polarimeter}

\author{R. Rezaei\inst{1}, C. Beck\inst{2,3}, W. Schmidt\inst{1}}
\titlerunning{Sunspot variations between 1999 and 2011 as observed with TIP}
\authorrunning{Rezaei, Beck \& Schmidt}  
   \institute{ Kiepenheuer-Institut f\"ur Sonnenphysik,
     Sch\"oneckstr. 6, D-79104 Freiburg, Germany.\\
     \email{[rrezaei, wolfgang]@kis.uni-freiburg.de}
    \and Instituto de Astrof\'{\i}sica de Canarias
     (CSIC), V{\'i}a Lact{\'e}a, E-38205 La Laguna, Tenerife, Spain
     \and Departamento de Astrof{\'i}sica, Universidad de La Laguna, E-38206 La Laguna, Tenerife, Spain\\
     \email{cbeck@iac.es}
   }

%

\abstract{} 
{We study the variation in the magnetic field strength and the umbral intensity of sunspots during the declining phase 
of the solar cycle no.~23 and in the beginning of cycle no.~24\rm.}
{We analyze a sample of 183 sunspots observed from 1999 until 2011 with the Tenerife Infrared Polarimeter\,(TIP) at the German 
Vacuum Tower Telescope (VTT)\rm. 
The magnetic field strength is derived from the Zeeman splitting of the Stokes-$V$ signal in one near-infrared spectral line, 
either \ion{Fe}{i}\,1564.8\,nm, \ion{Fe}{i}\,1089.6\,nm, or \ion{Si}{i}\,1082.7\,nm\rm. 
This avoids the effects of the unpolarized stray light from the field-free quiet Sun surroundings that can affect 
the splitting seen in Stokes-$I$ in the umbra. The minimum umbral continuum intensity and  umbral area are also measured.}
{We find that there is a systematic trend for sunspots in the late stage of the solar cycle no.~23 to be weaker, 
i.e., to have a smaller maximum magnetic field strength than those at the start of the cycle. 
The decrease in the field strength with time of about 
94\,G\,yr$^{-1}$ is well beyond the statistical fluctuations that would be expected because of the larger 
number of sunspots close to cycle maximum (14\,G\,yr$^{-1}$). 
In the same time interval, the continuum intensity of the umbra increases with a 
rate of 1.3\,($\pm\,0.4$)\% of \,I$_{c}$\,yr$^{-1}$, while the umbral area does not show any trend above the statistical variance. 
Sunspots in the new cycle no.~24 show higher field strengths and lower continuum intensities than those at the end 
of cycle no.~23, interrupting the trend.} 
{Sunspots have an intrinsically weaker field strength and brighter umbrae 
at the late stages of solar cycles compared to their initial stages, without any significant change in their area. 
The abrupt increase in field strength in sunspots of the new cycle suggests that the cyclic variations are dominating 
over any long-term trend that continues across cycles.  We find a slight decrease in field strength and 
an increase in intensity as a long-term trend across the cycles.}
\maketitle

\section{Introduction}
Morphological and physical studies of sunspots have been described for several decades \citep{bray_loughhead_64}. 
It was already known half a century ago that larger umbrae are darker 
and show higher magnetic field strength. 
\cite{albregtsen78} first reported a systematic variation in the umbral intensity in the infrared as a function 
of the solar cycle, with 
darker umbrae appearing in the early phase of the cycle \citep[see also][]{albregtsen_maltby_81}. 
\cite{adjabshir_koutchmy_83} attributed this variation to the change in the number of umbral dots 
within the solar cycle. This was dismissed by \cite{albergtson_etal_84}, who showed that the measured variations cannot be 
solely caused by umbral dots. Subsequent studies of the umbral intensity and the magnetic field 
strength established a relation between the continuum intensity and the magnetic field strength in sunspot
 umbrae \citep{kopp_rabin_92, martinez_vazquez_93}: the greater the field strength, the lower the intensity.  
While new high-resolution observations are still consistent with this empirical relation, no clear 
explanation has been provided \citep[see, however,][]{schuessler_etal_94}.

The variation in the field strength of sunspots in the last two cycles has been the subject of 
several studies \citep{livingston_02, penn_etal_03, norton_gilman_04, penn_macdonald_07,leonard_choudhary_08, schad_penn_10}. 
\cite{livingston_02} measured the separation of the $\sigma$-components of Stokes-$I$ of 
the \ion{Fe}{i} line at 1564.8\,nm, which is a Zeeman triplet with an effective Land\'e-factor of 3.0, 
and concluded that sunspots in cycle no.~23 were intrinsically weaker than those in cycle no.~22. 
A similar data set was also used in subsequent works \citep{livingston_etal_06, penn_livi_10}. 
These authors found that the field strength of sunspots was stronger at the beginning of cycle no.~22 
than at its end. Although the magnetic sensitivity of the infrared \ion{Fe}{i}\,1564.8\,nm line is 
high \citep{rueedi_etal_95}, the influences of scattered light and line blends in the intensity 
profiles on the measurement motivates one to employ more accurate methods to derive the field strength. 
It is not clear how dynamo theory can explain such cyclic dependency \citep[e.g.,][]{mathieu03}.

In this contribution, we used spectropolarimetric data in the \ion{Fe}{i} 1564.8\,nm line and 
other near-IR lines to evaluate the cyclic variation in the magnetic field strength of sunspots. 
We used full Stokes spectra to determine the magnetic field strength rather than solely using the intensity profiles. 
The advantage with respect to the stray light contamination is that the intensity in the quiet Sun (QS) surroundings 
is higher than in the umbra, whereas exactly the opposite relation holds for the polarization signal. 
Therefore intensity spectra in the umbra are contaminated by contributions from the QS, 
but the splitting of Stokes-$V$ is not. 
Another advantage of this sample compared to the previously mentioned 
works is that the majority of sunspots were observed with an active image correction; i.e., the spatial 
resolution of our data set is better than observations prior to the advent of correlation trackers or adaptive optics (AO).

Section \ref{sec:obs} describes the observations used. The data analysis methods are explained in Sect.~\ref{sec:dat}. 
Section \ref{sec:result} gives the results, together with numerical tests.  
The findings are discussed in Sect.~\ref{sec:disc}. Section \ref{sec:concl} 
provides the conclusions.

\section{Observations}\label{sec:obs}
We analyzed a sample of 231 sunspots observed from 1999 to 2011 with the Tenerife Infrared 
Polarimeter \citep[{TIP-I and TIP-II, respectively;}][]{martinez+etal1999, collados_etal07}. 
All observation were carried out at the German Vacuum Tower Telescope \citep[VTT,][]{vtt}. 
The observed maps usually cover a sunspot and the surrounding QS area. For some of the data taken with TIP-I 
using its small field of view (FOV) of about 30$^{\prime\prime}$, the sunspots' penumbra was not fully covered. 
Each data set consists of full Stokes profiles in magnetically sensitive infrared lines such 
as \ion{Fe}{i}\,1564.8\,m, \ion{Fe}{i}\,1089.6\,nm, or \ion{Si}{i}\,1082.7\,nm. Table \ref{tab:lines} summarizes 
the atomic parameters of the selected spectral lines. Out of the 231 maps, 99 were observed at 1.56\,$\mu$m and 84  
at 1.1\,$\mu$m. The remaining 48 maps mainly come from the earliest observations with TIP-I and were taken in some uncommon 
wavelength ranges, often covering molecular lines. These data could therefore not be used without considerable effort. 
We thus only selected those observations in which one of the three spectral lines listed in Table\,\ref{tab:lines} was recorded. 
That amounted to 183 full Stokes sunspot maps covering the descending phase of cycle no.~23 and the rise of cycle no.~24.

\begin{figure}
\resizebox{7.5cm}{!}{\includegraphics*{\filep  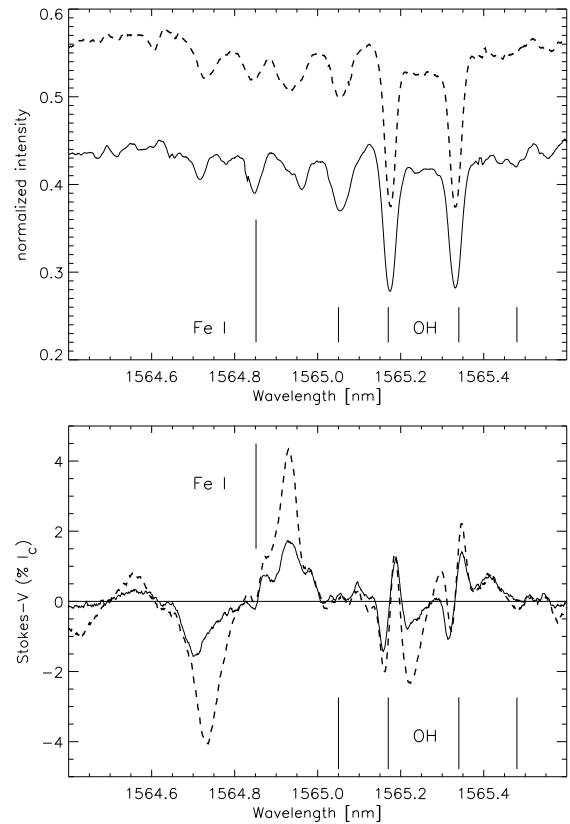}}
\caption{Sample of two umbral profiles. The solid (dashed) line corresponds to a sunspot at $\mu=0.58$ (0.76). 
The continuum intensity of the quiet Sun profiles corresponding to each of them 
is unity. The short vertical lines mark four OH spectral lines, while the long vertical line denotes the 
\ion{Fe}{i}\,1564.8\,nm line. \emph{Top:} Stokes-$I$, \emph{bottom:} Stokes-$V$.}
\label{fig:prof}
\end{figure}

The low temperature of umbrae allows certain molecules to form. As a result, the umbral profiles are usually contaminated with 
molecular blends \citep[e.g., the FTS atlas of umbral profiles of][]{umbral_atlas_fts}. Figure\,\ref{fig:prof} shows examples of 
umbral Stokes-$I$ and $V$ profiles at 1.56\,$\mu$m for two different heliocentric angles ($\mu=0.58$ and 0.76). 
The difference in the continuum level of the Stokes-$I$ profiles is real. 
As seen in this figure, there are several strong blends around the \ion{Fe}{i}\,1565.2\,nm line that corrupt not only 
the intensity profile but also the Stokes-$Q/U/V$ profiles of this line (bottom panel, Fig. \ref{fig:prof}). 
For this reason, we decided not to use the \ion{Fe}{i}\,1565.2\,nm line. The molecular blends are, however, 
useful for determining of atmospheric stratifications in inversions of umbral spectra \citep[e.g.,][]{mathew+etal2003}. 
There are several blends around the \ion{Fe}{i}\,1564.8\,nm line as well \cite[e.g.,][their Table\,1]{livingston_etal_06}. 
The main blend in this line is a CO line to the blue of the line center wavelength. 
It is, however, weaker than the OH blends near the \ion{Fe}{i}\,1565.2\,nm line. 

\begin{table}
\begin{center}
\caption{Atomic properties of the observed spectral lines~\citep{nave_etal94}.}
\label{tab:lines}
\begin{tabular}{c c c c c} 
\hline
Line & $\lambda$\,(nm) & Exc pot\,(eV) & $\log(gf)$ & $g$-effective\\\hline
Fe\,{\sc i} & 1564.852 & 5.426 & -0.669 & 3.00\\
Fe\,{\sc i} & 1089.630 & 3.071 & -2.845 & 1.50\\
Si\,{\sc i} & 1082.709 & 4.954 & 0.363 & 1.50\\
\hline
\end{tabular}
\end{center}
\end{table}

The spatial and spectral resolution, as well as the rms noise level, is different for each map. 
A careful analysis of the individual sunspots was done to take the diversity of the data into account. 
While the data until 2005 were recorded using a correlation tracker for real-time image 
correction \citep[CT,][]{ct_system}, 
the Kiepenheuer-Institute adaptive optics system \citep[KAOS,][]{luhe_etal_03} was 
utilized for all later data to improve the spatial resolution. 

\begin{figure}
\resizebox{8.0cm}{!}{\includegraphics*{\filep 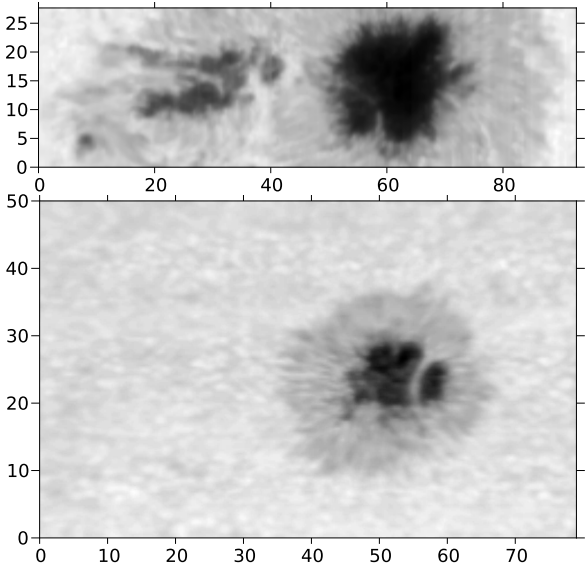}}
\caption{Example of TIP scans. \emph{Top:} sunspot on 24 May 2001, recorded in \ion{Fe}{i}\,1089.6\,nm with CT correction. 
\emph{Bottom:} sunspot on 17 July 2011, recorded in \ion{Si}{i}\,1082.7\,nm with AO image stabilization. Tick marks are in arcseconds.}
\label{fig:map}
\end{figure}

The old TIP-I camera \citep{martinez+etal1999} was used until mid 2006, providing a slit length 
of 27 arcsec at a spatial sampling of 0\farcs36 per pixel. The integration of the new TIP-II 
camera \citep[][]{collados_etal07} increased the slit length and at the same time doubled the spatial and spectral 
resolution because of the larger CCD chip and the smaller pixel size. 
Examples of data sets with the old and new TIP systems are shown in Fig. \ref{fig:map}. 
The top panel shows a TIP-I data set with CT correction, while the bottom panel 
shows a data set recorded with TIP-II  taken with AO correction. An overview of the data up to 2009 
is available online in the TIP 
archive\footnote{http://www3.kis.uni-freiburg.de/$\sim$cbeck/TIP$\_$archive/\\TIP$\_$archivemain.html. 
If you are interested in using the data, please contact mcv@iac.es.}. 
The header of each file contains information about, among other things, 
the location of the spot, scanning steps, and the integration time. 
This allowed us to retrieve the position and dimension of the sunspots. 
We used this information to calculate the area of the umbra and the sunspot as a whole.
The calibration of the data as well as the cross-talk correction was performed in the same manner as in \cite{schliche_collados02}.

\section{Data analysis}\label{sec:dat}
For each data set, an average quiet Sun profile was calculated using a region in the FOV outside of the penumbra. 
We used two of the spectral lines covered in each wavelength range to determine the spectral dispersion 
for each map. Then, the rms noise of the polarization signal, $\sigma$, was estimated in a continuum band. After that, 
the amount of separation between the two lobes of Stokes-$V$ was measured for all locations on the map with a maximum 
Stokes-$V$ amplitude larger than 12\,$\sigma$. ``Lobes'' were defined as local maxima in the polarization signal of
 Stokes-$V$ that a) were larger than the five wavelength pixels to the left and right and b) had an amplitude above the 
12-$\sigma$ level \citep[cf.~][]{cbeck_pdh,reza_etal_9}. This conservative threshold ensures that we only analyze 
clear antisymmetric Stokes-$V$ profiles with a drawback that we have some gaps in the maps in the neutral lines of 
Stokes-$V$ where the polarization signal strongly decreases. 
Because there are only a few data sets close to the limb ($\mu<$0.5), the neutral line  only passed through 
the umbra in a couple of cases. In particular, the 
missing values in the neutral line do not affect our results concerning the maximum magnetic field strength in umbrae. 
Whenever the profile exhibited more than two lobes, the two lobes most separated in wavelength were selected. 
Finally, we constructed maps of the magnetic field strength from the determined location of the two lobes in the 
Stokes-$V$ profiles and the spectral dispersion in the strong field approximation \citep{stix_book,landi_landolfi}:
\begin{equation}
 B = \frac{\Delta\lambda}{4.67\times10^{-12}\,\lambda^2\,g_{\mathrm{eff}}} \label{eq1}
\end{equation}
where $\lambda$ and $\Delta\lambda$ are in nm and $B$ is in Gauss. 
In the strong field limit, the separation of the two lobes is independent of the inclination angle, 
hence of the location of the spot on the disk. 
The signal amplitude is, however, affected by the inclination angle, but we do not use it here. 
In a plot of the maximum field strength {\it vs.} cosine of the heliocentric angle (not shown here), we did not find any significant trend. 
\begin{figure}
\resizebox{\hsize}{!}{\includegraphics*{\filep 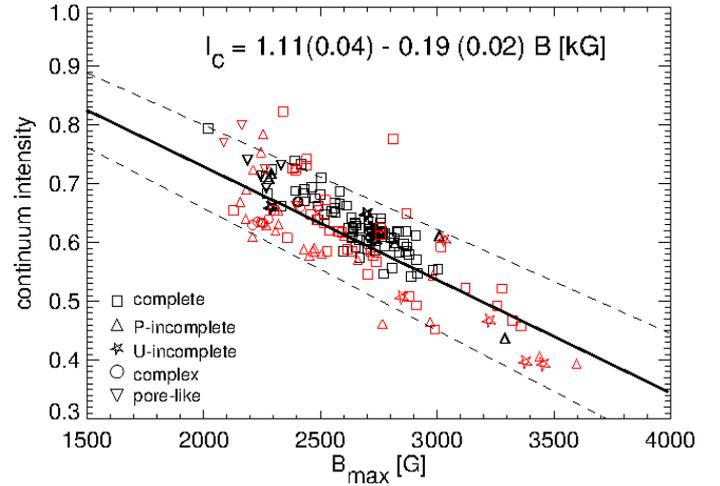}}
\caption{Variation in the minimum continuum intensity in the umbra as a function of maximum magnetic field strength.
The 1.1\,$\mu$m data was multiplied by a factor 1.3. Red and black symbols mark the 1.1 and 1.6\,$\mu$m data, respectively. 
P-incomplete and U-incomplete mark sunspots whose penumbra and umbra were not fully covered in the map. The solid line shows 
a least-squared fit, and the dashed lines indicate the 1$\sigma$ interval. Numbers in parenthesis denote the 1$\sigma$ uncertainty.}
\label{fig:cmin_bmax}
\end{figure}

The uncertainty of the estimated field strength was calculated using standard error propagation applied to Eq.\,(\ref{eq1}). 
The main source of uncertainty stems from the measurements of the separation of the lobes, whereas 
the effective Land\'e factor and the rest wavelength are known with high accuracy. 
We assumed an error of 0.1 pixels in the derivation of the position. 
This translates then into a ({\it random error}) $\sigma_{B}$ of about 20 to 60\,G, depending on the wavelength 
range and the dispersion of the spectra. The value is comparable to the formal error attributed to field strength 
in inversions of the \ion{Fe}{i} line at 1564.8\,nm \citep{beck_etal_07}. 
There are also systematic errors caused by gradients of the field strength with height in the atmosphere. 
With a typical measured vertical gradient of the magnetic field 
strength of $1-2$\,G\,km$^{-1}$\,\citep[e.g.,][]{schmidt_balthasar_94, westendrop_etal_2001}, 
this systematic error amounts to some $20-40$\,G difference in the measurements between $\mu=1$ and $\mu=0.5$.

Besides the maps of the magnetic field strength, we also determined the continuum intensity and the area of 
each sunspot. To define the umbra and penumbra, 
we produced at first an umbral mask using a variable threshold in continuum intensity. 
Using a fixed threshold did not produce satisfactory results because the umbral contrast 
varies with wavelength \citep[e.g.,][]{mattig1971,albregtsen_maltby_81,chapman_meyer_81,maltby_etal_86,tritschler+schmidt2002}. 
Using those (initial) masks, we constructed contours for the umbra and penumbra manually. 
Light bridges were excluded from the umbral masks. In some data sets, such as shown in the top panel of Fig.~\ref{fig:map}, 
the penumbra was not covered completely, so 
the penumbral area and the total sunspot area are underestimated in a few cases. 
The continuum intensities were not corrected for the limb-darkening effect. There was no significant trend 
in the minimum continuum intensity {\it vs.} the cosine of the heliocentric angle. 

A code was attributed to each spot describing its type in terms of complexity, completeness, and similar morphological 
characteristics  (see Sect.~\ref{sec:empirical}). 
In the case of complex active regions (e.g., the top panel of Fig.~\ref{fig:map}), only the largest spot in the map was selected.
\begin{figure}
\resizebox{8cm}{!}{\includegraphics*{\filep 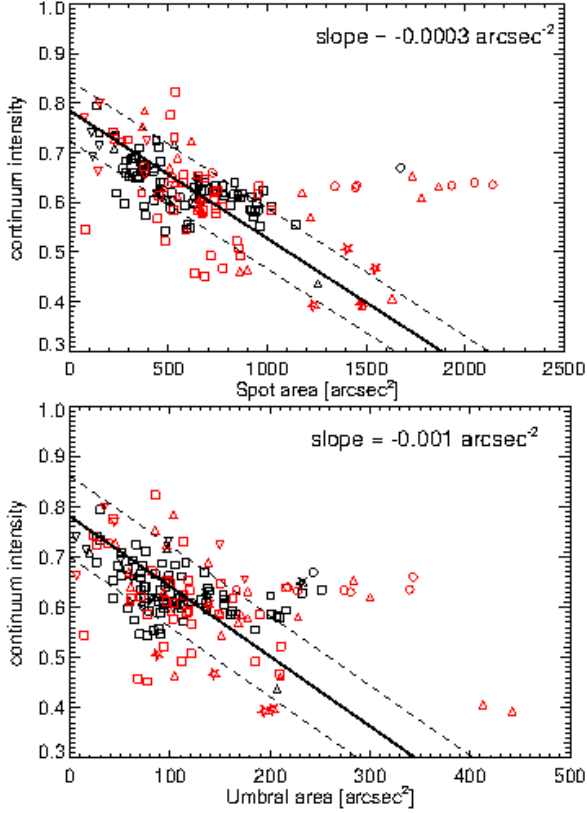}}
\caption{Variation in the minimum continuum intensity in the umbra as a function of area. \emph{Top:} total sunspot area, 
\emph{bottom:} umbral area. The solid lines indicate  
a bisector least-squared fit to all data points expect complex sunspots (circles). 
The dashed lines show the 1$\sigma$ interval. Symbols are like Fig. \ref{fig:cmin_bmax}.}
\label{fig:area_cmin}
\end{figure}
\begin{figure}
\resizebox{8cm}{!}{\includegraphics*{\filep 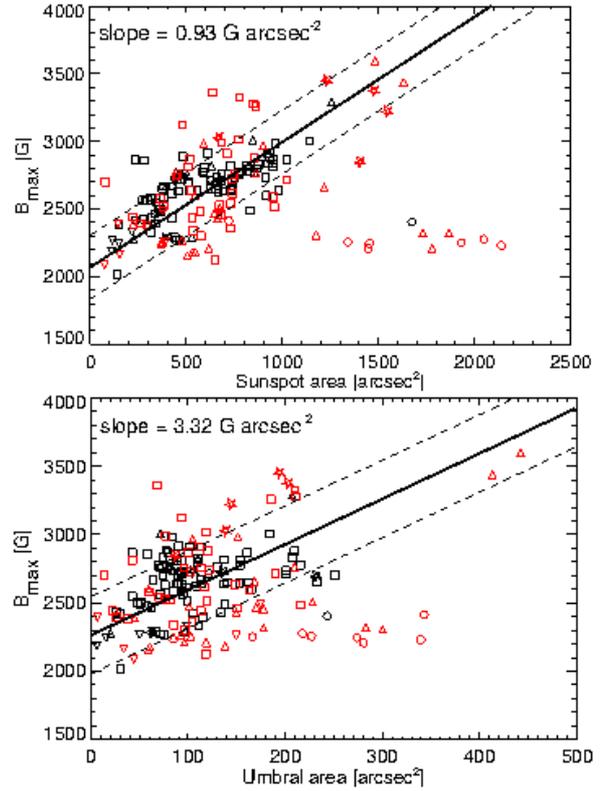}}
\caption{Variation in the maximum magnetic field strength in the umbra as a function of area. 
\emph{Top:} total sunspot area, \emph{bottom:} umbral area. Symbols are like Fig \ref{fig:cmin_bmax}. 
Linear fits are similar to Fig.~\ref{fig:area_cmin}.}
\label{fig:area_bmax}
\end{figure}

The continuum intensity sample is not fully homogeneous because we have the continuum intensity in two different wavelength windows 
(1.1 and 1.6\,$\mu$m). As mentioned above, sunspots have higher contrasts at shorter wavelengths. 
We multiplied the continuum intensity of sunspots at 1.1\,$\mu$m with a factor of 1.3 to have the same scale as in 1.6\,$\mu$m, 
both in scatter plots and histograms (Sect.~\ref{sec:result}). 
This coefficient of relative umbral intensities of 1.3 corresponds to an umbral temperature of about 4200\,K. 
The intensities in each wavelength range were normalized to the intensity in the QS surroundings of 
the sunspot. For a temperature of about 6000\,K in the QS, the ratio of continuum intensities in the Planck curve 
is $I_{\rm QS}(\lambda=1100\,{\rm nm})/I_{\rm QS}(\lambda=1565\,{\rm nm}) \sim 2.77$; i.e., the intensity at the 
shorter wavelength has to be multiplied by 2.77 to be normalized to $I_{\rm QS}(\lambda=1565\,{\rm nm})$. 
This yields a relative umbral intensity 
$I_{\rm umbra}(\lambda=1100\,{\rm nm})/I_{\rm QS}(\lambda=1565\,{\rm nm}) = 1 / 1.3 \cdot 2.77 \sim 2.13$
, when the umbral intensity at 1565\,nm is set to unity. A ratio of 2.13 in the intensities 
$I(\lambda=1100\,{\rm nm})/I(\lambda=1565\,{\rm nm})$ is obtained in the Planck curve for $T= 4165$\,K. 
The value fits to previous determinations of umbral temperatures 
\citep[e.g.,][]{maltby_etal_86,collados_etal_94,mathew+etal2003,solanki03r,cuberes_etal_05,beck2008}. 

\section{Results}\label{sec:result}
\subsection{Empirical relations}\label{sec:empirical}
Figure \ref{fig:cmin_bmax} shows the scatter plot of the minimum umbral intensity {\it vs.} the maximum field strength, 
B$_{\mathrm{max}}$. 
By ``minimum'' we mean the average of the ten pixels with the lowest intensity. 
Similarly by B$_{\mathrm{max}}$, we mean the average of the ten pixels with the largest field strength. 
These pixels partly overlap those with the lowest intensity. 
Different symbols in the plot indicate morphological details of each sunspot. Triangles denote sunspots whose penumbra touches 
the map border, although they have a complete umbra. Circles mark sunspots in complex active region. 
This group of sunspots sometimes have extended or incomplete penumbrae. 
\begin{figure*}
\resizebox{\hsize}{!}{\includegraphics*{\filep 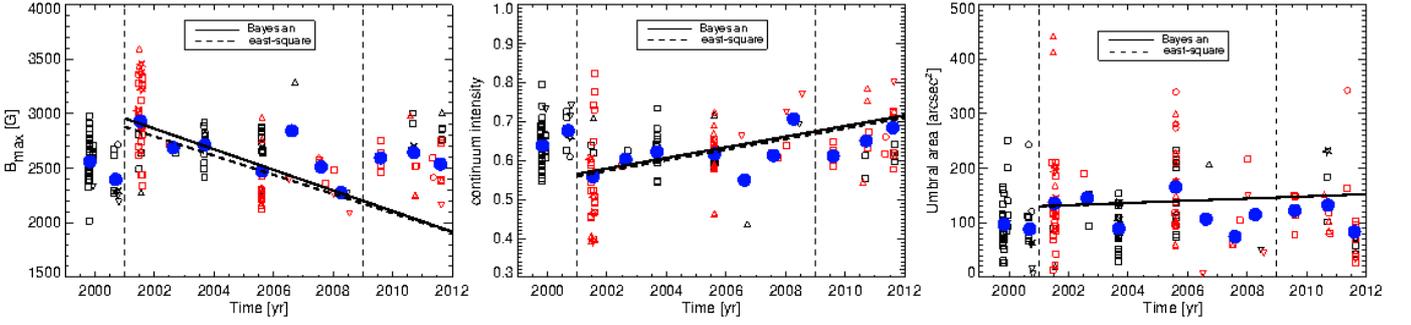}}
\caption{Variation in the maximum magnetic field strength (left), minimum continuum intensity (midddle), and umbral area (right) 
{\it vs.} time. Symbols show the individual sunspots while the blue filled circles mark the annual averages. 
The linear fits was performed for individual data points between the two vertical lines. 
The dashed vertical lines mark maximum and minimum of cycle 23. Symbols are like in Fig \ref{fig:cmin_bmax}. }
\label{fig:tempo}
\end{figure*}

As a result, a triangle, for instance, does not exclude the spot belonging to a complex active region. 
Reverse triangles mark those objects that are either a pore or more like a protospot: they do not have a fully fledged penumbra. 
In rare cases, the map does not fully cover the umbra. These objects are marked with asterisks. As seen in Fig.~\ref{fig:cmin_bmax}, 
there is a tight correlation between the continuum intensity and the field strength. 
The classical or Pearson correlation (C$_{\mathrm{p}}$), as well as the nonparametric 
Spearman correlation coefficients (C$_{\mathrm{s}}$), were calculated \citep{press_etal_92}. 
We find correlation coefficients of C$_{\mathrm{p}}$\,=\,0.80 and C$_{\mathrm{s}}$\,=\,0.76. 
The solid line in Fig.~\ref{fig:cmin_bmax} is a standard least-squared fit with an slope 
of -0.19 ($\pm\,1.5\times\,10^{-2}$) of $I_c$\,kG$^{-1}$ and an offset of 1.11($\pm\,0.04$) of $I_c$. 
The offset of 11\,\%  (solid line, Fig.~\ref{fig:cmin_bmax}) above unit continuum intensity 
for $B\equiv 0$\,G matches stray light estimates in the umbra well, 
where stray light levels between 8\,\% and 15\,\% were found for similar data 
\citep[e.g.,][]{rezaei+etal2007,beck2008,beck+etal2011}. 

Figure \ref{fig:area_cmin} shows the variation in minimum continuum intensity {\it vs.} area. 
In this figure as well as Fig.~\ref{fig:area_bmax}, we do not show the error bars because they are usually smaller 
than the size of the symbols. 
There are a few points that are off the main trend.  
A close inspection showed that those sunspots belong to complex active regions and have extended penumbrae. 
If we omit their extended penumbrae, their areas shrink significantly. 
As seen in Fig.~\ref{fig:area_cmin}, there is an anticorrelation between the minimum continuum intensity and umbral (sunspot) area. 
The correlation coefficients are C$_{\mathrm{p}}$\,=\,-0.43 and C$_{\mathrm{s}}$\,=\,-0.40 
(C$_{\mathrm{p}}$\,=\,-0.48 and C$_{\mathrm{s}}$\,=\,-0.57), respectively. 
This is consistent with earlier findings: larger umbrae are intrinsically cooler.  
The linear fits in Figs.~\ref{fig:area_cmin} and \ref{fig:area_bmax} are bisector fits. 
That means we assume that both variables are independent \citep{linefit}. 
For these four fits, the complex sunspots (circles) were excluded. 
A standard least-square fit results in shallower slopes. 
Assuming a linear relation between the B$_{\mathrm{max}}$ and minimum intensity with the sunspot/umbral area 
is equivalent to the relation used by other authors \citep[e.g., ][]{schad_penn_10}, who fit a quadratic function between 
these quantities and the umbral radius.

Figure \ref{fig:area_bmax} shows the variation in the B$_{\mathrm{max}}$ {\it vs.} area.  
The trend is clearer with the sunspot area than the umbral area. 
We find a positive correlation 
between the maximum field strength and the umbral (sunspot) area (Fig. \ref{fig:area_bmax}). For B$_{\mathrm{max}}$ {\it vs.} umbral (sunspot) 
area, we find C$_{\mathrm{p}}$\,=\,0.26 and C$_{\mathrm{s}}$\,=\,0.20 (C$_{\mathrm{p}}$\,=\,0.25 and C$_{\mathrm{s}}$\,=\,0.39), respectively. 
If we skip the complex sunspots with the largest area, the correlation coefficient increases significantly. 
We created two random vectors of the same size as our TIP sample 
and calculated their correlations. After repeating this experiment for 1000 times, 
the standard deviation both for the Pearson and Spearman correlations were $\sigma_{\mathrm{C}}=0.07$, so  
the found correlation coefficients are well above the distribution of purely random numbers. 

Part of the scatter in Figs.~\ref{fig:area_cmin} and \ref{fig:area_bmax} 
is real because there is a diversity in properties of simple and complex sunspots, 
as well as differences between fully developed and forming sunspots \citep{collados_etal_94}.
As discussed in Sect.\ref{sec:dat}, the separation of the two lobes of Stokes-$V$ in the strong-field regime is not 
a function of the inclination angle nor the heliocentric angle.

\subsection{Temporal evolution}\label{sec:bay}
Figure \ref{fig:tempo} shows the temporal evolution of the maximum field strength, the minimum continuum 
intensity, and the umbral area. 
Symbols display individual umbrae while the blue circles mark annual averages. 
The two dashed vertical lines at 2001 and 2009 denote the maximum and minimum of the 
solar cycle no.~23 (defined from the sunspot number). 
A decrease in the magnetic field strength is seen in the diagram (left panel). 
%
We examined correlations to check for any systematic trend of B$_{\mathrm{max}}$ with the solar cycle. To this extent, we selected 
the time span between the two dashed lines in Fig.\,\ref{fig:tempo}. The correlation coefficient for the magnetic field 
strength is about C$_{\mathrm{p}}$\,=\,-0.57 and C$_{\mathrm{s}}$\,=\,-0.43 for the Pearson and Spearman definitions, respectively. 
This indicates the significance of the decrease in magnetic field strength in the declining phase of solar cycle no.~23. 

We used a Bayesian method \citep[e.g.][]{gregory_loredo_92, bayes} to fit a straight line to the individual data points between 
2001 and 2009. To this end, a Markov Chain Monte Carlo (MCMC) method was utilized with one million iterations \citep{pymc}. 
This resulted in a linear fit with an offset of 2915\,(\,$\pm$\,30)\,G and a slope of -94\,(\,$\pm$\,7)\,G\,yr$^{-1}$. 
The solid line in Fig.\,\ref{fig:tempo} shows the best-fit line. 
The dashed line shows a standard least-square linear fit whose slope is -88\,($\pm\,14$)\,G\,yr$^{-1}$. 
The extension of the solid line (Fig.\,\ref{fig:tempo}) to 2012 demonstrates that the decrease in the 
field strength does not continue in cycle no.~24.
If we use the range of 2000 to 2009 instead, the slopes drop to -59\,(\,$\pm$\,8) G\,yr$^{-1}$  and 
-56\,($\pm\,14$)\,G\,yr$^{-1}$ for the Bayesian and least-square methods, respectively.

As seen in the middle panel of Fig.\,\ref{fig:tempo}, 
there is a weak tendency in sunspots such that the umbra is brighter at the end of the cycle. 
Both linear fits have a common slope of 1.3\,($\pm\,0.4$)\,\%\,I$_{\mathrm{c}}$\,yr$^{-1}$.  
From the relation between the umbral intensity and  B$_{\mathrm{max}}$ (Sect.~\ref{sec:empirical}), 
one expects to find an increase in the continuum intensity because we find a decrease in the field strength. 
The amount of decrease of B$_{\mathrm{max}}$ during the declining phase is about 750\,G, leading to  a predicted increase in 
the minimum intensity of some 14\,\%\,of I$_{\rm{c}}$. 
As a result, the expected slope for the temporal evolution of minimum intensity 
based on the empirical relation between B$_{\mathrm{max}}$ and intensity (Fig.~\ref{fig:cmin_bmax}) is 
about 1.8\,\%\,yr$^{-1}$, consistent with the linear fit within the 1$\sigma$ uncertainty range.  

The correlation coefficient of the umbral area {\it vs.} time (right panel, Fig.\,\ref{fig:tempo}) 
in the same time interval is C$_{\mathrm{p}}$\,=\,0.05 
and C$_{\mathrm{s}}$\,=\,0.12 for the Pearson and Spearman correlation coefficients, respectively. 
It stays insignificant for the deprojected area of the umbra. 
The corresponding coefficients for the sunspot area are C$_{\mathrm{p}}$\,=\,-0.06 and C$_{\mathrm{s}}$\,=\,-0.03, respectively, 
so such correlations are not significant (recall the limit of $\sigma_{\mathrm{C}}$=0.07 
for purely random variables, Sect.~\ref{sec:empirical}). 
That means that within the scatter and statistics of our data, there is 
no systematic change in the umbral size with the solar cycle. 

\subsection{Distribution of parameters}
\begin{figure}
\resizebox{8cm}{!}{\includegraphics*{\filep 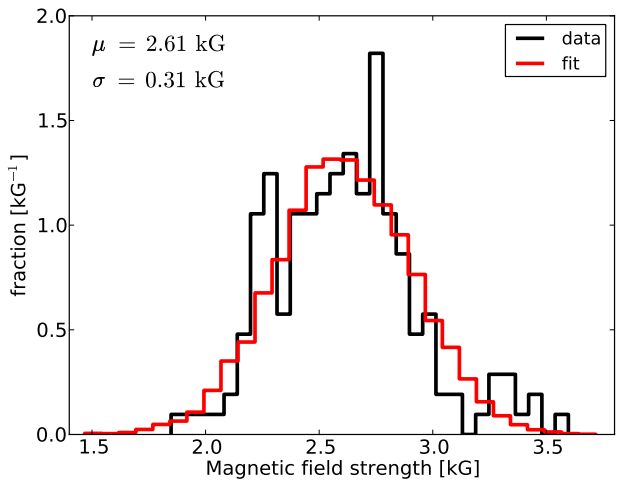}}
\caption{Probability density function of the maximum magnetic field strength. 
The red curve is the best-fit normal distribution to the data using a Bayesian method.}
\label{fig:bhist}
\end{figure}

The probability density function (PDF) of the maximum field strength is shown in Fig. \ref{fig:bhist}. 
A Kolmogorov-Smirnov test was applied to find the best PDF for the data \citep{press_etal_92}. 
This test allows one to reject that the observed and fit distributions are significantly different. 
A normal distribution with an average field strength of 2.6\,kG and a variance of 0.3\,kG reproduces the 
observed PDF of the field strength well. There is no umbra with a magnetic field strength below about 1.8\,kG. 
The assumed normal distribution thus exceeds the observed PDF for field strengths below that value. 
In the limit of extremely large field strengths above about 3\,kG, the best-fit PDF also does not  
matches the observed one perfectly. This will presumably be caused by the poor statistics of those few umbrae 
with a maximum field strength above 3\,kG in the observations, and the selection of the largest sunspot in complex active regions.
\subsection{Numerical experiment}\label{sec:num_exp}
\begin{figure}
\resizebox{\hsize}{!}{\includegraphics*{\filep 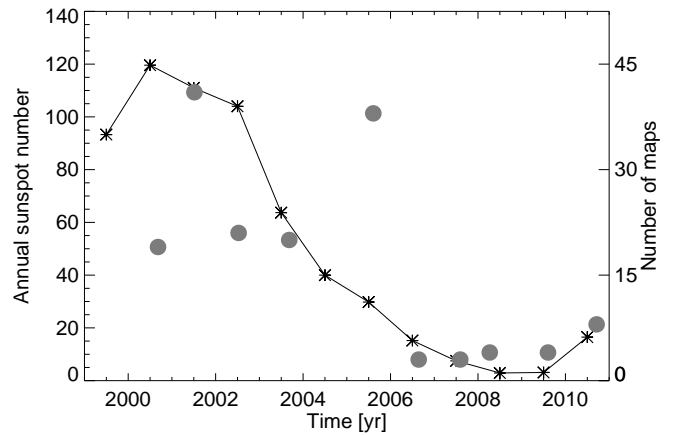}}
\caption{Yearly sunspot number. The solid line shows the SIDC annual sunspot number. The filled circles mark the 
number of TIP maps in each year.}
\label{fig:cycle}
\end{figure}

To investigate the influence of noise and systematic effects on the derived slope of B$_{\mathrm{max}}$ with the 
phase of the solar cycle, we performed three numerical tests. This allowed us to evaluate the 
amount and importance of spurious signals in our results. In particular, we tried to evaluate the significance 
of the trend of field strength with the phase of the solar cycle (-94\,G\,yr$^{-1}$).
\paragraph{First experiment:} Assume that the maximum magnetic field strength of sunspots has a constant distribution at 
any moment of a solar cycle, i.e., the PDF of the maximum field strength, B$_{\mathrm{max}}$, is the same for each year 
and is given by Fig.~\ref{fig:bhist}. The number of sunspots close to a solar maximum is about 100 times more 
than at the solar minimum (cf.~Fig.~\ref{fig:cycle}). Therefore, the statistics of sunspots around the time of maximum activity 
is better and will sample the PDF correctly, whereas close to minimum activity strong deviations from the true PDF may occur 
because of the sparse sample. This can lead to an apparent temporal trend in field strength caused purely by the variation of 
the sample size, where both a decrease or an increase in the field strength can result.
To reject the hypothesis that the measured systematic variation of -94\,G\,yr$^{-1}$ is caused only by the large 
number of sunspots close to the maximum of a cycle and the low number close to the minimum (Fig.~\ref{fig:cycle}), 
we performed the following test. 

Figure \ref{fig:cycle} shows the yearly sunspot number from the Solar Influences Data Center (SIDC) during cycle no.~23 
and the early years of cycle no.~24 \citep{sidc}. 
For each year from 1999 through 2009, we created a sample of B$_{\mathrm{max}}$, according to the observed distribution 
of field strengths (Fig. \ref{fig:bhist}). The size of the sample population was set to the annual sunspot number,   
making the PDFs of the the field strength in individual years roughly identical within the statistical fluctuations. 
We then analyzed this synthetic temporal sample in the same manner as the real data.
Using both the Bayesian (Sect. \ref{sec:bay}) and the least-square methods, we fit a straight line to measure 
any change of the mean of B$_{\mathrm{max}}$. The test was repeated 100 times. 
The average and rms of the resulting slopes were zero and 11\,G\,yr$^{-1}$, respectively. 
The two methods resulted in identical average and rms slopes. The sample of 100 slopes had a 
maximum and minimum of +23 and -26\,G\,yr$^{-1}$, respectively. We note that, if one plots all such artificial 
solar cycle curves, the width of the region covering the 100 curves increases toward the end of the cycle as expected 
because of the poorer statistics. 
That means the upper boundary for statistical fluctuations of less than 30\,G\,yr$^{-1}$ as the sole reason for 
the observed systematic trend is significantly smaller than the slope we have derived from the observations. 
\paragraph{Second experiment:} 
Although the first test showed that the trend is real, it is not clear how far it is influenced by the statistical 
fluctuations. We performed a second numerical experiment to elaborate on this issue. Like the first test, 
we created artificial samples of B$_{\mathrm{max}}$. The sample size for each year was as in the first test. 
The PDF of the  B$_{\mathrm{max}}$ was, however, different for each year in the sense that the mean value of 
the PDF was decreased by -94\,G\,yr$^{-1}$. Because of the lower number of sunspots at the end of 
the cycle, it is then possible to get a slope that is too flat or too steep compared to the actually imposed 
slope. The number of repetitions and the analysis method  were identical to the first experiment. 
The average and rms of the slopes derived in the second experiment were -94 and 12 \,G\,yr$^{-1}$, 
respectively, both for the Bayesian and least-square methods. The maximum and minimum slopes of the sample were 
-68 and -130\,G\,yr$^{-1}$, respectively. That means that with a complete sample the measured slope can be off 
by about 36\,G\,yr$^{-1}$ in a worst case scenario.
\paragraph{Third experiment:} 
The second experiment showed that, if there is a real trend in B$_{\mathrm{max}}$ with the phase of the solar cycle, 
it is measurable despite the different statistics of sunspots at the maximum and minimum of the cycle. 
However, we do not have {\it all} of the sunspots in our sample, only those that were observed with TIP. 
This can introduce  a significantly larger uncertainty in the derived slope compared to the second experiment. 
Therefore, we repeated the second numerical test with one change: instead of 
using the annual sunspot number of SIDC, we used the statistics of the TIP sample. 
The resulting slopes then spanned a range between -130  and -66\,G\,yr$^{-1}$ with an average of -84 
and a standard deviation of 9\,G\,yr$^{-1}$. This means the statistics of the TIP sample has about 
the same amount of uncertainty as the complete sample with a caveat that it underestimates the correct trend 
by 10\,G\,yr$^{-1}$, well within the 1$\sigma$ range of 14\,G\,yr$^{-1}$. 
This ensures us that the slope of -94\,G\,yr$^{-1}$ is significant and measurable with the statistics of our sample. 
\section{Discussion}\label{sec:disc}
\subsection{Tracers of the solar cycle}
 The number of sunspots and sunspot groups are the most common indicators for studying the solar cycle 
\citep[][ and references therein]{hathaway_etal_2002}. Nevertheless, other parameters with more straightforward 
physical meanings show similar, if not better traces of cyclic variations. These parameters, such as the magnetic 
field strength of sunspots, usually have a caveat: their systematic recording started only in the early 20th century. 
In other words, they cannot compete with the 400-year record of sunspot numbers. 
Recent investigations of the sunspot numbers also indicate that large sunspots correlate better with chromospheric 
and coronal indicators of the solar cycle such as the 10.7\,cm radio flux \citep{kilcik_etal_2011}.

\subsection{Continuum intensity vs. magnetic field strength}
Our results reproduce earlier observations regarding the general behavior of the 
maximum magnetic field and minimum continuum intensity of sunspots: 
the greater the field strength, the lower the continuum intensity. 
Our Fig.~\ref{fig:cmin_bmax} compares reasonably well with Fig.\,2 of \cite{livingston_02}. Similar curves of 
the relation between intensity and field strength for this and other wavelength ranges are given in literature 
\citep[e.g.,][]{martinez_vazquez_93,lites_elmore_etal93,keppen_martinez_96, stanchfield+etal1997,norton_gilman_04,leonard_choudhary_08, wesolowski_etal_08,schad_penn_10}, 
where those of \citet{gurman_house81} and \citet{livingston_02} 
match ours closest because of the pronounced linear behavior. 
In some of the works the relation between intensity and field strength was derived inside a single sunspot, not for a 
sunspot sample, sometimes also yielding clearly nonlinear relations.
\subsection{Temporal evolution of B$_{\mathrm{max}}$}
Solar cycle no.~23 was weaker than cycle no.~22 \citep[e.g.,][]{ata_2006}. 
\cite{livingston_02} report that sunspots in cycle no.~23 had an excess of smaller and brighter umbrae 
compared to cycle no.~22. Using 1.56\,$\mu$\,m Kitt Peak data, \cite{penn_livi_10} suggest that the strengths 
of sunspots has steadily decreased since the last solar maximum. Although our finding (Fig. \ref{fig:tempo}) 
in the declining phase of cycle no.~23 is consistent with them, we do find an increase in  B$_{\mathrm{max}}$ 
since the beginning of the solar cycle no.~24. 
As seen in the lefthand panel of Fig. \ref{fig:tempo}, the average field strength of the 
years 2010 and later are well above the linear trend that indicates the extrapolation of the decrease during the declining 
phase of cycle no.~23. That is in contrast to Fig.~1 of \cite{penn_livi_10}, who predict a continuous decrease 
in the sunspot field strength and consequently on the solar magnetic activity across cycles. 
We note, however, that the field strength at the start of cycle no.~24 seems to be slightly lower than at the same stage 
in cycle no.~23, and the continuum intensity correspondingly slightly higher. This will have to be confirmed by additional 
data in the following years.

The weakening rate of the magnetic field strength in the descending phase of cycle no.~23 in the TIP data is 
about -94\,($\pm\,14$)\,G\,yr$^{-1}$. 
\cite{penn_livi_06} reported  -52\,G\,yr$^{-1}$.
The size of our sample is smaller than that of \cite{livingston_02}, as well as \cite{penn_livi_06}. 
However, the precision of the measurements, and the calibration accuracy is superior to the mentioned references 
because of using vector-polarimetric data instead of intensity profiles.

\cite{watson_etal2010} used  Soho magnetograms and analyzed more than 30,000 sunspots. 
They reported a systematic decrease of -70\,G\,yr$^{-1}$ in the declining phase of the cycle no.~23, in agreement with our results. 
These results should be taken with care because the Soho/MDI instrument cannot accurately measure the magnetic field 
in sunspots because of line saturation effects \citep{berger_lites_2003}.
\cite{pevtsov_etal_2011} compiled observations of seven (former Soviet) solar observatories. 
They report a rate of -119\,G\,yr$^{-1}$ for cycle no.~23, which is the strongest rate in 
the four cycles they studied. Despite some (systematic) uncertainty in part of their data around 1965, 
they did not find any significant secular trend in field strength over four and a half solar cycles.
\subsection{Temporal evolution of umbral intensity}
\cite{albregtsen78} and \cite{albergtson_etal_84} found that the amount of variation in the umbral intensity during a 
cycle is about 10\,\% \citep[see also,][]{maltby92}. Our results are therefore consistent with their analysis.
\cite{mathew_etal_2007} used Soho data and did not find any significant enhancement in the umbral 
intensity as a function of the solar cycle phase. In contrast, \cite{penn_macdonald_07} find a cycle-dependent oscillation 
of the umbral intensity.  Our data show an increase in the umbral continuum intensity with the phase of the solar cycle. 
We find a slope  of 1.3\,($\pm\,0.4$)\,\% I$_{\mathrm{c}}$($1.56\,\mu$\,m)\,\%\,yr$^{-1}$ in the TIP data of cycle 
no.~23. Our results are consistent with the observations of \cite{albergtson_etal_84},  who reported an increase of 
1.2 and 1.9\,\% per year in cycles no.~20 and 21, as well as with \cite{penn_livi_06} who find 
an increase of 1.9\,\% per year from 2000 to 2006. 
\subsection{Umbral area}
\cite{brants_zwaan_82} found that larger umbrae have a higher field strength, similar to \cite{kopp_rabin_92} 
and \cite{martinez_vazquez_93}. We find a similar trend in our results (Fig.~\ref{fig:area_bmax}). Recently, \cite{watson_etal2010} 
report a dependence of the area of the largest sunspots on the solar cycle. 
 This, however, does not imply any trend in the size of individual sunspots. 
Like \cite{penn_macdonald_07}, we did not find any systematic trend between the phase of the solar cycle and the sunspot size. 
These observations therefore hint at a constant distribution of the sunspot area, e.g., a log-normal distribution with a 
fixed slope \citep{bogdan88}. The larger number of sunspots close to the maximum of a cycle naturally 
increases the number of large sunspot groups. Because these large sunspots are located at the tail of the distribution, 
they are prone to larger statistical fluctuations compared to other quantities resulting from all sunspots such as B$_{\mathrm{max}}$. 
The lack of a systematic change of sunspot area is still puzzling because we measure a difference of some 700\,G 
between the average B$_{\mathrm{max}}$ in the maximum and minimum of cycle no.~23, comparable to the 600\,G 
reported by \cite{penn_livi_06}. Taking Fig.~\ref{fig:area_bmax} at face value shows the relation between field strength 
and area, a rate of -28\,arcsec$^2$\,yr$^{-1}$ would be expected. A least-square fit to the average umbral areas per year resulted, 
however, only in a slope of 2\,($\pm$\,4)arcsec$^2$\,yr$^{-1}$. 
It was already mentioned by \cite{brants_zwaan_82} that the scatter between B$_{\mathrm{max}}$ and umbral size is partly intrinsic. 
\cite{penn_macdonald_07} suggest that the relation between the B$_{\mathrm{max}}$ and umbral area might vary during a solar cycle. 
Therefore, it remains an open question to verify whether the diversity of physical conditions 
in different umbrae vary with the solar cycle in a sense that it prevents a cyclic trend of umbral size. 

\cite{mathew_etal_2007} report a clear trend between the umbral intensity and the umbral 
size in Soho data \citep[see also][]{wesolowski_etal_08}. Although the umbral contrast is lower at 1.56\,$\mu$m, 
there is  a similar trend in our data as well. As seen in Fig.~\ref{fig:area_cmin}, larger sunspots have a 
lower continuum intensity than smaller sunspots, in agreement with previous studies 
\citep{brants_zwaan_82, kopp_rabin_92}.

\subsection{Implications for the solar irradiance}
An excess or lack of large dark umbrae changes the total solar irradiance \citep[TSI, e.g., ][]{tsi_1990}. 
Our finding thus also has some implications for the TSI variations in solar cycle no.~24. There are long term variations in 
the solar activity and irradiance \citep[e.g.,][]{tap_etal_2007, froh_2009, shapiro_etal_2011}. In particular, because the Sun 
apparently had an unusually active period in the last couple of decades \citep{usoskin_etal_2003, bonev_etal_2004}, 
we cannot completely rule out the existence of a long term trend. 
However, in spite of suggestions that cycle no.~24 and the preceding activity minimum are very different from previous ones \citep{tripathy+etal2010,jain+etal2011,tap_etal_2011}, the influence of a long term 
trend seems to be less than the cycle-dependent pattern on the timescale of a solar cycle.
\section{Conclusion}\label{sec:concl}
A sample of  sunspot maps observed with the Tenerife Infrared Polarimeter was analyzed to elaborate on a possible dependence of 
the magnetic field strength of sunspots on the phase of the solar cycle. The vector-polarimetric data enabled us to use full Stokes 
measurements to estimate the variation in B$_{\mathrm{max}}$ rather than 
only using intensity profiles as used in case of Kitt Peak data. 
We found a significant decrease in the measured magnetic field strength from the maximum to the minimum of the sunspot 
cycle no.~23 with a rate of -94 ($\pm\,14$) G\,yr$^{-1}$. 
At the same time, the minimum umbral continuum intensity increases by 1.3\,($\pm\,0.4$)\% of $I_c$\,yr$^{-1}$, while there 
is no significant trend of the umbral area with the phase of the solar cycle.
The magnetic field strength of sunspots increased at the beginning of solar cycle no.~24. This indicates that the 
observed trend of B$_{\mathrm{max}}$ is likely to be a cyclic pattern rather than a long-term evolution. 
Comparing the field strength in the early stages of cycle nos.~23 and 24, we find a slight reduction in B$_{\mathrm{max}}$ 
in the new cycle that, however, will need to be confirmed by additional data in the future.

\begin{acknowledgements}
The German VTT is operated by the
Kiepenheuer-Institut f\"ur Sonnenphysik at the Spanish Observatorio del Teide of the Instituto de Astrof\'{\i}sica de
Canarias (IAC). 
We are grateful to Manuel Collados (IAC) for providing the data. 
We also thank Juan M. Borrero for reading the manuscript. 
R.R. acknowledges fruitful discussions at the workshop on 
``Filamentary Structure and Dynamics of Solar Magnetic Fields'', as well as 
``Extracting Information from spectropolarimetric observations: comparison of inversion codes'' at the ISSI in Bern. 
C.B. acknowledges partial support by the Spanish Ministry of Science and Innovation through projects AYA 2007-63881 and AYA 2010-18029.
\end{acknowledgements}

\bibliographystyle{aa}
\bibliography{rezabib_nn}
\end{document}